\documentstyle[preprint,aps,floats,epsfig,subeqn]{revtex}

\newcommand{\be}{\begin{equation}}
\newcommand{\ee}{\end{equation}}
\newcommand{\bea}{\begin{eqnarray}}
\newcommand{\eea}{\end{eqnarray}}
\newcommand{\bml}{\begin{mathletters}}
\newcommand{\eml}{\end{mathletters}}

\begin{document}

\tighten

\preprint{DCPT-02/43}
\draft

%\twocolumn[\hsize\textwidth\columnwidth\hsize\csname @twocolumnfalse\endcsname

%%%%%%%%%%%%%%%%%%%%%%%%%%%%%%%%%%%%%%%%%%%%%%%%%%%%%%%%%%%%%%%%%%%%%%%%%%

%\wideabs{                       % Uncomment this line for two-column output

\title{Bound monopoles in Brans-Dicke theory}
\renewcommand{\thefootnote}{\fnsymbol{footnote}}

\author{Betti Hartmann\footnote{Betti.Hartmann@durham.ac.uk}}
\address{Department of Mathematical Sciences, University
of Durham, Durham DH1 3LE, U.K.}
\date{\today}
\setlength{\footnotesep}{0.5\footnotesep}

\maketitle
%%%%%%%%%%%%%%%%%%%%%%%%%%%%%%%%%%%%%%%%%%%%%%%%%%%%%%%%%%%%%%%%%%%%%%%%%%
\begin{abstract}
We consider axially symmetric
SU(2) Yang-Mills-Higgs (YMH) multimonopoles in Brans-Dicke theory
for winding number $n > 1$. In analogy to the spherically
symmetric $n=1$ solutions, we find that the axially symmetric solutions
exist for higher values of the gravitational coupling than
in the pure Einstein gravity case. For large values of
the gravitational coupling, the solutions collapse to form a
black hole which outside the horizon can be described by an extremal
Reissner-Nordstr\"om solution. Similarly as in the pure Einstein gravity case,
like-charged monopoles reside in an attractive phase in a limited domain
of parameter space. However, we find that the strength of
attraction is decreasing for decreasing Brans-Dicke parameter $\omega$.
\end{abstract}


%%%%%%%%%%%%%%%%%%%%%%%%%%%%%%%%%%%%%%%%%%%%%%%%%%%%%%%%%%%%%%%%%%%%%%%%%%
%\newpage
\renewcommand{\thefootnote}{\arabic{footnote}}
\section{Introduction}

Topological defects \cite{vilenkin} are believed to have formed during
phase transitions in the early universe and are related to
spontaneously broken symmetries. Depending on 
the topology of the vacuum manifold,  $d$-dimensional objects form.
The $d=0,1,2$ defect is the magnetic monopole, the string and the domain wall,
respectively. 
Two basic types of topological defects have been considered in the literature~:
a) defects which are described by theories with a broken global symmetry
and b) defects which are
described by theories with a broken local, i.e. gauge symmetry
such as the Nielsen-Olesen vortex \cite{nielsen} and the 'tHooft-Polyakov 
monopole \cite{thooft}. While the latter exhibit particle-like properties 
such as finite energy and a well defined core, the former have 
divergent energies. In turn, due to their long-range behaviour, global
defects have much stronger gravitational effects 
\cite{barriola,harari,cohen,ruth} than their local counterparts 
\cite{vilenkin2,bfm}.

The local magnetic monopole results from a spontaneous symmetry breaking of
an SU(2) gauge symmetry down to a U(1) gauge symmetry. The mapping of spatial
infinity to the vacuum manifold is characterized by the winding number $n$. 
The spherically symmetric $n=1$ solution is the famous 'tHooft-Polyakov monopole
\cite{thooft}. Since this was proved to be the unique spherically symmetric solution
\cite{bogo}, the construction of multimonopoles longed for
an axially symmetric Ansatz which was introduced in \cite{rebbi}.\

 In flat space like-charged monopoles
are non-interacting \cite{bps} in the limit of vanishing
Higgs boson mass. In this  so-called Bogomol'nyi-Prasad-Sommerfield
(BPS) \cite{ps,bogo} limit, the attraction of the long-range Higgs field
exactly compensates the repulsion caused by the long-range
U(1) field. These configurations
saturate a lower energy bound, the so-called Bogomol'nyi bound  such that
their energy is proportional to the winding number $n$.
For finite Higgs mass, the monopoles are  repelling \cite{kkt}
since now the Higgs field is exponentially decaying.

Gravitating monopoles have been studied in the framework of General Relativity
\cite{bfm,hkk}. Interestingly, it was found that 
gravitating monopoles
can - for specific choices of the gravitational constant
and the Higgs self-coupling constant - reside in an attractive phase 
in the sense that their energy per
winding number is decreasing with increasing $n$ \cite{hkk}. 

Among alternative gravity theories,
the scalar-tensor theory introduced by Brans and Dicke \cite{bd} is one of the
most popular. Since Mach's principle \cite{mach} leads to problems within the
framework of General Relativity, Brans and Dicke introduced a scalar field
which plays an analog role than the inverse of Newton's constant and is coupled 
to the system by the so-called Brans-Dicke (BD) parameter $\omega$. However,
the interest in BD theory in recent years was mainly motivated by the fact
that the scalar-tensor part of
the low energy effective action of superstring 
theory resembles BD theory \cite{string}.

Global topological defects in Brans-Dicke like theories have been studied extensively
\cite{global}. In \cite{tmt} spherically symmetric solutions of SU(2) Brans-Dicke-Yang-Mills-Higgs (BDYMH)
theory in Schwarzschild-like coordinates have been constructed. 
This includes both the globally regular monopole as well
as the corresponding black hole solutions. 

In this paper, we construct axially symmetric monopole solutions of SU(2) BDYMH theory. We give the
Lagrangian, the Ansatz and the boundary conditions in Section II and present our numerical
results in Section III. The conclusions are summarized in Section IV. 

\section{SU(2) Brans-Dicke-Yang-Mills-Higgs (BDYMH) theory}
In \cite{tmt} it was argued that for conformally invariant
fields such as the Yang-Mills fields, the Lagrangian in the Einstein frame equals that
in the BD frame. In the following, all solutions - unless otherwise stated - are 
those obtained in the Einstein frame. 
\subsection{The Lagrangian}
The Lagrangian $\tilde{{\cal L}}$ of Brans-Dicke theory reads \cite{bd}~:
\begin{equation}
\label{lagbd}
\tilde{{\cal L}}=\frac{1}{16\pi G}\left(\tilde{\Psi} \tilde{R} -\frac{\omega}{\tilde{\Psi}}
\tilde{\partial}_{\mu}\tilde{\Psi}\tilde{\partial}^{\mu}\tilde{\Psi}\right)
+\tilde{{\cal L}}_m
\end{equation}
$G$ denotes Newton's constant and  $\omega > -3/2$ is the Brans-Dicke parameter.
For $|\omega|\rightarrow \infty$ standard Einstein gravity is recovered.
A conformal transformation to the Einstein frame yields 
the Lagrangian ${\cal L}$  \cite{tmt}~:
\begin{equation}
\label{action}
{\cal L}=\frac{1}{16\pi G}R-\frac{1}{2}\partial_{\mu}\Psi\partial^{\mu}\Psi+
\frac{2\omega+3}{2\omega+4}{\cal L}_m
\end{equation}
with $\Psi$ given by the BD scalar field $\tilde{\Psi}$~:
\begin{equation}
\label{BDfield}
\Psi=\frac{1}{\gamma\sqrt{8\pi
G}}\left(\ln(\tilde{\Psi})-\ln(\frac{2\omega+4}{2\omega+3})\right) \ , \
\gamma=(\omega+\frac{3}{2})^{-1/2}  \ . \
\end{equation}
The matter Lagrangian ${\cal L}_m$ is given by~:
\begin{equation}
\label{lag}
{\cal L}_m=-\frac{1}{4}F_{\mu\nu}^{a}F^{\mu\nu,a}
-\frac{1}{2}e^{-\gamma\sqrt{8\pi G}\Psi}D_{\mu}\Phi^{a} 
D^{\mu}\Phi^{a}-e^{-2\gamma\sqrt{8\pi G}\Psi}V(\Phi^{a})
\ ,   \end{equation}
with Higgs potential
\begin{equation}
V(\Phi^{a})=\frac{\lambda}{4}(\Phi^{a}\Phi^{a}-\eta^2)^2
\ , \end{equation}
the non-abelian field strength tensor
\begin{equation}
F_{\mu\nu}^{a}=\partial_{\mu}A_{\nu}^{a}-\partial_{\nu}A_{\mu}^{a}+
e\varepsilon_{abc}A_{\mu}^{b}A_{\nu}^{c}
\ , \end{equation}
and the covariant derivative of the Higgs field in the adjoint 
representation ($a=1, 2, 3$)
\begin{equation}
D_{\mu}\Phi^{a}=\partial_{\mu}\Phi^{a}+
e\varepsilon_{abc}A_{\mu}^{b}\Phi^{c}
\ . \end{equation}
$\lambda$ and $\eta$ are
the Higgs field's self-coupling constant and vacuum expectation
value, respectively and $e$ denotes the
gauge coupling constant. 
The prefactors of the covariant derivative and
the Higgs potential in (\ref{lag}) result from the conformal transformation
from the BD to the Einstein frame~:
\begin{equation}
g_{\mu\nu}=\frac{2\omega+3}{2\omega+4}\tilde{\Psi} \tilde{g}_{\mu\nu}
\end{equation}
Note that the Lagrangian (\ref{action}) resembles that
of Einstein-Yang-Mills-Higgs-dilaton theory \cite{bh,bhk} with a specific coupling
of the scalar field. 
\subsection{Axially symmetric Ansatz}
The axially symmetric Ansatz for the metric in isotropic coordinates reads \cite{bk2}~:
\begin{equation}
ds^2=
  - f dt^2 +  \frac{m}{f} \left( dr^2+ r^2d\theta^2 \right)
           +  \frac{l}{f} r^2\sin^2\theta d\varphi^2
\ .  \end{equation}
For the gauge fields we choose the purely magnetic Ansatz \cite{rebbi}~:
\begin{equation}
{A_t}^a=0 \ , \ \ \  {A_{r}}^a=\frac{H_1}{er}{v_{\varphi}}^a 
\ , \end{equation}
\begin{equation}
{A_{\theta}}^a= \frac{1-H_2}{e} {v_{\varphi}}^a
\ , \ \ \ \ 
{A_{\varphi}}^a=- \frac{n}{e}\sin\theta \left(H_3{v_{r}}^a+
(1-H_4) {v_{\theta}}^a \right)
\ . \end{equation}
while for the Higgs field, the Ansatz reads \cite{rebbi}:~
\begin{equation}
{\Phi}^a=\eta (\Phi_1 {v_{r}}^a+\Phi_2 {v_{\theta}}^a)
\ . \end{equation}
The vectors $\vec{v}_{r}$,$\vec{v}_{\theta}$ and $\vec{v}_{\varphi}$
are given by:
\begin{eqnarray}
\vec{v}_{r}      &=& 
(\sin \theta \cos n \varphi, \sin \theta \sin n \varphi, \cos \theta)
\ , \nonumber \\
\vec{v}_{\theta} &=& 
(\cos \theta \cos n \varphi, \cos \theta \sin n \varphi,-\sin \theta)
\ , \nonumber \\
\vec{v}_{\varphi}   &=& (-\sin n \varphi, \cos n \varphi,0) 
\ .\label{rtp} \end{eqnarray} 
Here the winding number $n$ enters the Ansatz for the fields.
Since we are considering static axially symmetric solutions,
the function $f$, $l$, $m$, $H_1$, $H_2$, $H_3$, $H_4$, $\Phi_1$, $\Phi_2$
 and $\Psi$ depend only on $r$
and $\theta$. The spherically symmetric Ansatz for the construction
of $n=1$ solutions in isotropic coordinates is recovered if
the dependence on $\theta$ is dropped
 and additionally $m=l$, $H_1=H_3=\Phi_2=0$, $H_2=H_4$.\

We introduce the
following dimensionless variable $x$ and the dimensionless field $\psi$~:
\begin{equation}
\label{scale}
x=e\eta r \ , \ \ \ \psi=\sqrt{8\pi G}\Psi
\end{equation}
With this rescaling, the set of differential equations arising 
from the variation 
\begin{equation}
\delta S=\delta \left(\int {\cal L} \sqrt{-g} d^4 \ x\right)=0
\end{equation}
depends only on $\omega$ and the following 
dimensionless coupling constants~:
\begin{equation}
\alpha=\sqrt{4\pi} \frac{M_W}{eM_{Pl}}=\sqrt{4\pi G}\eta \ , \ 
\beta=\frac{M_{H}}{\sqrt{2}M_W}=\frac{\sqrt{\lambda}}{e}
\end{equation}
where $M_W=e\eta$ is the gauge boson mass, $M_H=\sqrt{2\lambda}\eta$ is the
Higgs boson mass and $M_{Pl}=G^{-1}$ is the Planck mass.

The energy $E(n)$ of the solutions for this choice of Ansatz is given in terms of
the derivative of the metric function $f$ at infinity~:
\begin{equation}
E(n)=\frac{1}{2\alpha^2}\lim_{x\to\infty} x^2\partial_x f \ .
\end{equation}
\subsection{Boundary conditions}
We have to impose $10$ conditions on each of the
boundaries $x=0$, $x=\infty$, $\theta=0$ and $\theta=\frac{\pi}{2}$ \cite{theta}
to solve the set of $10$ second order partial differential equations. Regularity at the origin
requires~:
\begin{equation}
\partial_x f(0,\theta)=\partial_x l(0,\theta)=
\partial_x m(0,\theta)=0,\ \ \partial_x \psi(0,\theta)=0
\end{equation}
\begin{equation}
H_i(0,\theta)=0,\ i=1,3 ,\ \ H_i(0,\theta)=1,\ i=2,4,\ \
\Phi_i(0,\theta)=0,\ i=1,2 
\end{equation}
At infinity, the requirement for finite energy and asymptotically
flat solutions leads to the boundary conditions:
\begin{equation}
\label{bcinf}
f(\infty,\theta)=l(\infty,\theta)=m(\infty,\theta)=1,\ \ \psi(\infty,\theta)=0
\end{equation}
\begin{equation}
H_i(\infty,\theta)=0,\ i=1,2,3,4 ,\ \ \Phi_1(\infty,\theta)=1,\ \
\Phi_2(\infty,\theta)=0 
\end{equation}
In order to obtain the right symmetry for
the solutions, we set on  the $z$-axis as well as on the $\rho$-axis 
($\theta=0$ and $\theta=\pi/2$, respectively)~:
\begin{equation} 
H_1=H_3=\Phi_2=0
\end{equation}
and
\begin{equation}
\partial_\theta f=\partial_\theta m=\partial_\theta l 
=\partial_\theta H_2=\partial_\theta H_4=
\partial_\theta \Phi_1=\partial_\theta \psi=0
\end{equation}
\section{Numerical results}
We have solved the set of partial differential equations numerically. First, we studied
the behaviour of the solutions for fixed $\omega$, $\beta$ and increasing
gravitational coupling $\alpha$. In FIG.~1,
we show the energy per winding number $E/n$ for $\beta=0$, $n=2$ 
and three different
values of $\omega$. $\omega=\infty$ corresponds to the pure Einstein gravity limit,
studied in \cite{hkk} and for $\alpha=0$, the BPS multimonopoles are recovered
with energy per winding number (in our rescaled variables) equal to 
unity. For $\omega < \infty$ and $\alpha=0$, the energy per winding number
is that of the BPS multimonopoles but rescaled by the prefactor
$(2\omega+3)/(2\omega+4)$ of the matter Lagrangian ${\cal L}_m$ (see (\ref{action})).
The reason for this is that for $\alpha=0$, the Brans-Dicke function
and the metric functions are trivial $\psi\equiv 0$ and $f=m=l\equiv 1$, respectively.
Thus the matter Lagrangian 
${\cal L}_m$, which is proportional to the energy density, is that of the flat
space Yang-Mills-Higgs multimonopoles. This reasoning is, of course, also true for $\beta\neq 0$. Thus~:
\begin{equation}
E(\alpha=0,n,\omega,\beta)=\frac{2\omega+3}{2\omega+4}
E(\alpha=0,n,\omega=\infty,\beta) \ . \
\end{equation}
For $\alpha\rightarrow\alpha_{max}$ the branch of
Brans-Dicke multimonopoles bifurcates with the branch of extremal
Reissner-Nordstr\"om (RN) solutions. The extremal RN solution
in the model studied here is given by~:
\begin{equation}
\label{RNf}
f(x)=\left(\frac{x}{x+\hat{\alpha} n}\right)^2 \ , \ \ \ \hat{\alpha}=\alpha\sqrt{
\frac{2\omega+3}{2\omega+4}} 
\end{equation}
and
\begin{equation}
\label{RNrest}
H_i(x)=\Phi_2(x)=0 \ , \ i=1, 2, 3, 4  \ , \  \Phi_1(x)=1 \ , \  \psi(x)=0 
\end{equation}
Note that the horizon of the extremal RN solution
in isotropic coordinates is located at $x=x_h=0$. The functions
of the limiting solution thus correspond to those of the RN solution
on the full interval $x~\epsilon~ [0:\infty[$.  The solution (\ref{RNf})
and (\ref{RNrest}) has magnetic charge $P$ and energy $E(n)$ depending on the
Brans-Dicke parameter $\omega$~:
\begin{equation}
P=n\alpha\sqrt{
\frac{2\omega+3}{2\omega+4}} \ , \ \ \ \  E(n)=P/\alpha^2
\end{equation}
In FIG.~1, we demonstrate 
that for all chosen values of $\omega$, the branch of
globally regular monopole solutions bifurcates with the corresponding branch of
extremal RN solutions. Moreover, we 
observe that $\alpha_{max}(\omega)$, the maximal value of $\alpha$ for which globally regular monopole
solutions exist, is increasing  drastically for decreasing $\omega$ (see also FIG.~3). 
We find, e.g. for $\beta=0$, $n=2$~:
\begin{equation}
\alpha_{max}(\infty)\approx 1.49 \ , \ \ \alpha_{max}(0)\approx 1.89\ , \ \  
\alpha_{max}(-1)\approx 2.80
\end{equation} 
This agrees with the results in \cite{tmt}, where it was found that
for the $n=1$ solutions the ratio $\alpha_{max}(0)/\alpha_{max}(\infty)\approx 1.3$. 

In FIG.~2, we demonstrate the behaviour of the metric function $f$
in the limit $\alpha\rightarrow\alpha_{max}$ for $\omega=0$, $n=2$ and $\beta=0$.
$f$ is shown as function of the compactified coordinate
$z=x/(1+x)$ \cite{z}. For increasing $\alpha$, the value of the metric function
at the origin, $f(0)$, decreases to the RN value $f_{RN}(0)=0$. 
Moreover, the angle dependence  diminishes indicating that 
the limiting solution is the spherically symmetric RN solution.
For $\alpha\rightarrow\alpha_{max}\approx 1.89$, $f$ approaches
the function given by (\ref{RNf}).

For $\omega=\infty$ and $\alpha=0$, like-charged monopoles are either
non-interacting (in the limit of vanishing Higgs 
boson mass) or repelling. When gravity comes into play,  it is possible
for the multimonopoles to reside in an attractive phase \cite{hkk}.
In the BPS limit ($\beta=0$), this is not suprising, for $\beta\neq 0$ however it is apparent
that for small values of $\beta < \hat{\beta}(n)$ gravity is able to overcome the repulsion of the
long-range magnetic field. In \cite{hkk} it was found that
$\hat{\beta}(n=2)\approx 0.21$. Nevertheless, the extension of the attractive phase
is limited by the fact that the solutions exist only up to an $n$- and $\beta$-
dependent  maximal value of the gravitational coupling $\alpha$.
We find that 
like Einstein gravity Brans-Dicke gravity is 
able to overcome the repulsion between like-charged monopoles for sufficiently
high values of the gravitational coupling $\alpha$.
In FIG.~3, we show $\alpha_{eq}(\beta,\omega)$, the value of $\alpha$ for which
the mass of the $n=1$ monopole and the mass per winding number of the $n=2$ 
multimonopole equal one another, as function of $\omega$ for two different
values of $\beta$.  For both values of $\beta=0.1$ and $\beta=0.2$, $\alpha_{eq}$ is increasing
for decreasing $\omega$ from its value at $\omega=\infty$. Equally,
the value $\alpha_{max}$ is increasing with decreasing $\omega$. The multimonopoles
reside in an attractive phase in the parameter space above the $\alpha_{eq}$-curve
and below the corresponding $\alpha_{max}$-curve. Thus in analogy
to the $\omega=\infty$ limit Brans-Dicke gravity is
able to overcome the repulsion of the long-range U(1) field
for $\beta=0.1$ and $\beta=0.2$ and sufficiently high values of $\alpha$. 
Since $\alpha_{max}$ is increasing drastically for small $\omega$, 
the extention of the attractive phase for small $\omega$ is much bigger 
than in the $\omega=\infty$ limit. However, we find that the order
of magnitude of the value $\Delta=E(n=1)-E(n)/n$, which is an indicator
for the strength of attraction between like-charged monopoles
doesn't change significantly within the whole domain of  attraction
in the $\alpha$-$\omega$-plane. This can be seen from the table below where
we show $\Delta(\omega)=E(n=1,\omega)-E(n=2,\omega)/2$ for
$\alpha_{max}(n=2,\omega)/k$, $k=2,3,4$~:
\begin{center}
Table 1
\medskip

\begin{tabular}{|l|c|c|c|}
\hline \hline
$k$ & $2$ & $3$ & $4$ \\
\hline
$\Delta(\omega=\infty) $ & 0.0051 & 0.0024 & 0.0022  \\
\hline 
$\Delta(\omega=-1) $  & 0.0052 & 0.0033 & 0.0018  \\
\hline   
\end{tabular}
\end{center}

In FIG.~4, we show the difference between the
energy of the $n=1$ monopole and the $n$ multimonopoles, $\Delta=E(n=1)-E(n)/n$,
as function of $\omega$ for $\alpha=1$, $\beta=0$ and $n=2,3,4$, respectively.

$\Delta$ stays nearly constant for a large range of $\omega$ and decreases
for decreasing $\omega$. For $\omega \rightarrow \omega_0= -3/2$, $\Delta$ decreases 
to zero which is due to the fact that the prefactor $(2\omega+3)/(2\omega+4)$ is equal
to zero for $\omega=-3/2$ and thus the mass of the solutions itself vanishes.
It can be clearly deduced from FIG.~4 that the strength of attraction between
like-charged monopolos is smaller in  Brans-Dicke
theory than in pure tensor gravity theory. This can be compared with the
results in \cite{bh_dila} and \cite{bh_volkov} for two different
Einstein-Yang-Mills-Higgs-dilaton (EYMHD) models. EYMHD theory - like Brans-Dicke
theory - constitutes a theory of gravity in which the metric tensor has
a scalar companion, in the case of EYMHD, the dilaton. While in \cite{bh_dila}
the standard coupling of the dilaton in $4$ dimensions was studied,
the model studied in \cite{bh_volkov} arose from
dimensional reduction of an Einstein-Yang-Mills system in $(4+1)$ dimensions
\cite{volkov}. In both models the dependence
of the strength of attraction on the dilaton coupling was studied. It was found
that the value $\Delta$ was increasing for increasing dilaton coupling
with $\alpha$ and the dilaton coupling not 
too close to their maximal values. 
Since in the model studied here, 
$\sqrt{8\pi G}\gamma(\omega)$ (see (\ref{BDfield})) can be interpreted as the analog 
of a dilaton coupling and is increasing for decreasing $\omega$ and fixed $G$ (i.e. fixed $\alpha$),
the strength of attraction is decreasing for increasing "dilaton" coupling.

Apparently, the strength of attraction is increasing for increasing $n$
which suggests that large clumps of monopoles might be possible. However,
FIG.~4 also gives a hint  that monopoles in such clumps will be more 
bound in pure Einstein gravity ($\omega=\infty$) than in Brans-Dicke gravity for
equal gravitational coupling $\alpha$.

\section{Conclusions and Summary}
We have studied axially symmetric multimonopoles with $n > 1$ in Brans-Dicke
gravity. We find that for a maximal value of the gravitational coupling $\alpha$
these solutions collapse to form an abelian
black hole which outside the horizon is described
by an extremal Reissner-Nordstr\"om (RN) solution with trivial
Brans-Dicke scalar field. The RN solution has mass and
magnetic charge depending on the Brans-Dicke parameter $\omega$. This behaviour
can be compared to that in Einstein-Yang-Mills-Higgs-dilaton (EYMHD) 
theory \cite{bhk,bh_dila}.
When either the gravitational coupling or the dilaton coupling reaches its maximal
value, the branch of monopole 
solutions bifurcates with the branch of extremal Einstein-Maxwell-dilaton
(EMD) solutions. These have a naked singularity and
a non-trivial scalar dilaton field.

In analogy to the $n=1$ solutions, the maximal value of the gravitational
coupling up to where the globally regular monopole solutions exist, is increasing
for decreasing Brans-Dicke parameter. Equally, $\alpha_{eq}$, the value
of $\alpha$ where the mass of the $n=1$ monopole and the energy per winding number
of the $n=2$ multimonopole equal one another, is increasing for
decreasing $\omega$. At the same time, $\alpha_{max}$, the maximal value of
the gravitational coupling $\alpha$ up to where globally regular multimonopole
solutions exist, is increasing. Remarkable is that the rate of increase
for decreasing $\omega$ is much bigger for $\alpha_{max}$ than for $\alpha_{eq}$.
Thus, the extension of the attractive phase is much bigger for small $\omega$
than for $\omega=\infty$. We conclude that  at comparable
values of the gravitational coupling $\alpha$, the Brans-Dicke monopoles
are less bound than the pure Einstein gravity monopoles, and though the former
can exist to much bigger values of the gravitational coupling than
the latter, the strength of attraction is of the same order of magnitude
for all values of $\alpha$ and $\omega$ for which  bound multimonopoles
exist. These phenomena can be explained by noticing that in the gravitational field equations
always the combination $\alpha_{eff}:= \alpha\sqrt{(2\omega + 3)/(2\omega +4)}$
appears. This can be interpreted as an "effective" gravitational constant.
Clearly, for $\alpha$ fixed and $\omega$ decreasing from infinity, this expresssion
is decreasing. This explains why the value $\alpha_{max}$ up to where the Brans-Dicke
multimonopoles exist is increasing for $\omega$ decreasing and also
why at fixed $\alpha$, the strength of attraction is decreasing for 
decreasing $\omega$. Only at comparable values of $\alpha$ for different $\omega$
(e. g. at the $k$-th part of $\alpha_{max}(\omega)$) is the strength of attraction
of the same order of magnitude in Einstein theory and Brans-Dicke theory, respectively. 

In \cite{tmt} the corresponding spherically symmetric black hole solutions
were studied with emphasis on their thermodynamic properties.
Axially symmetric black hole solutions in Einstein-Yang-Mills-Higgs theory
have been studied recently \cite{hkk_bh} in the context of the so-called
"Isolated horizon framework" \cite{acs}. It would be interesting to analyse
in which sense the corresponding black hole solutions of the model
studied in this paper fulfill the predictions of this framework.
\\
\\
{\bf Acknowledgements}
This work was supported by the EPSRC.

\newpage
\begin{figure}\centering\epsfysize=13cm
\mbox{\epsffile{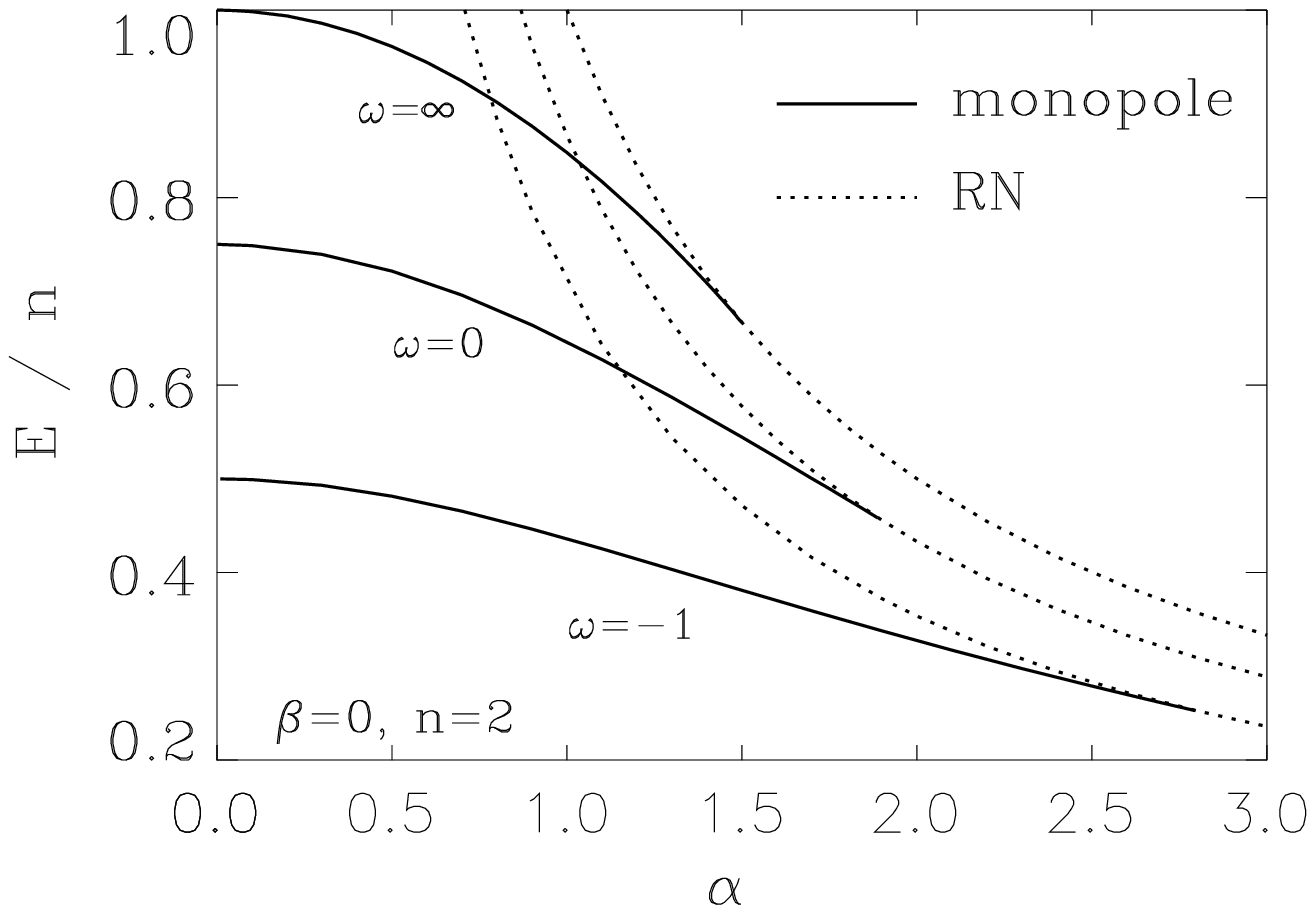}}
\caption{The energy per winding number $E/n$ of the axially symmetric
$n=2$ Brans-Dicke
monopole (solid) is shown for 
$\beta=0$, $\omega=\infty$, $0$ and $-1$ as function of $\alpha$.
Also shown is the energy per winding number
of the corresponding Reissner-Nordstr\"om (RN) solution (dotted). }
\end{figure}
\newpage
\begin{figure}\centering\epsfysize=13cm
\mbox{\epsffile{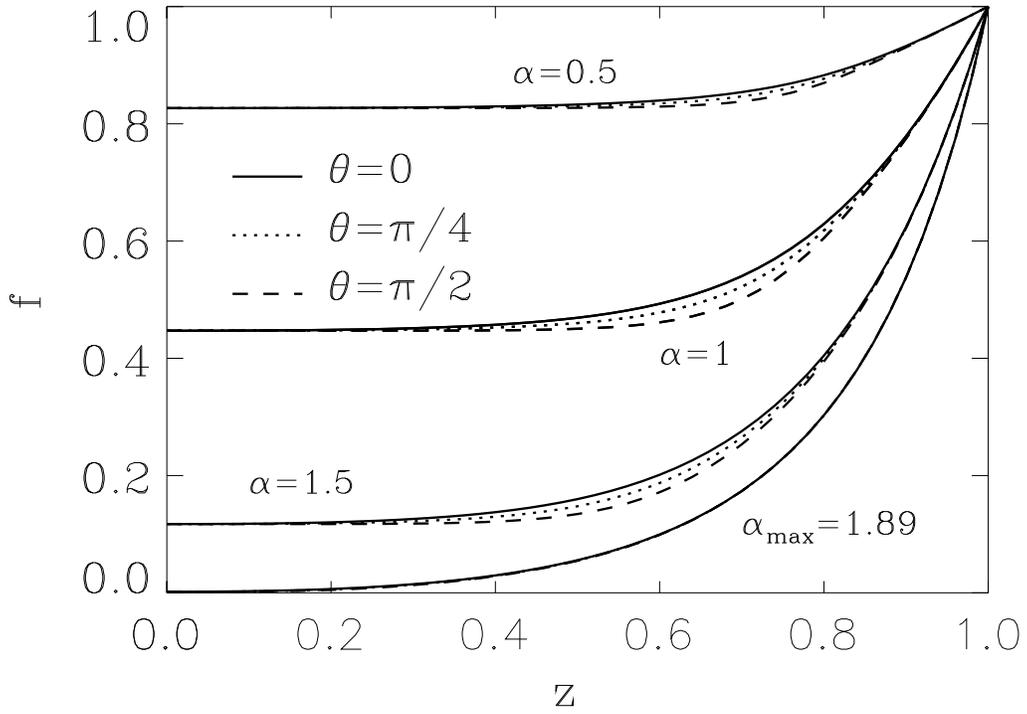}}
\caption{The metric function $f$ is shown as function of the
compactified coordinate $z=x/(1+x)$ for $n=2$, $\omega=0$, $\beta=0$ and
several values of $\alpha$ including $\alpha=\alpha_{max}\approx 1.89$.
 }
\end{figure}

\newpage
\begin{figure}\centering\epsfysize=13cm
\mbox{\epsffile{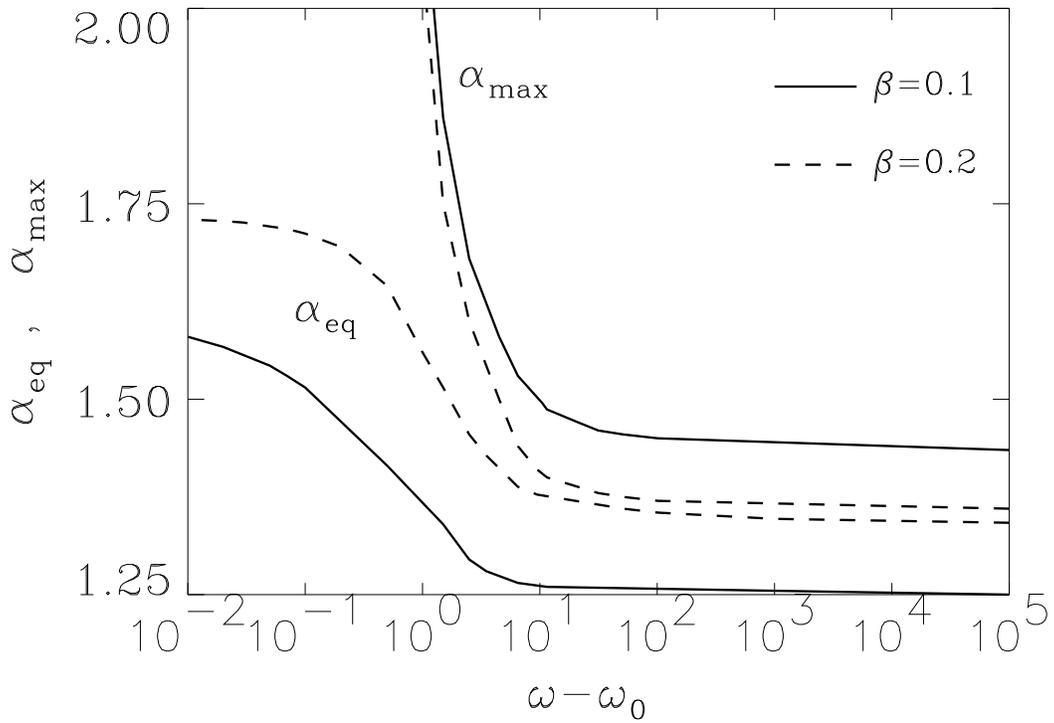}}
\caption{The value $\alpha_{eq}$, where the mass per winding number
of the $n=2$ multimonopole and the $n=1$ monopole equal one another is shown
for $\beta=0.1$ and $\beta=0.2$ as function of $\omega-\omega_{0}$, 
$\omega_{0}=-\frac{3}{2}$. Also shown is $\alpha_{max}$ for $n=2$ and
the same values of $\beta$. Note that the values of $\alpha_{max}$ increase 
drastically with decreasing $\omega$, e.g. $\alpha_{max}(\beta=0.1,\omega=-1)
\approx 2.7$ and 
$\alpha_{max}(\beta=0.2,\omega=-1)\approx 2.6$. }
\end{figure}
\newpage
\begin{figure}\centering\epsfysize=13cm
\mbox{\epsffile{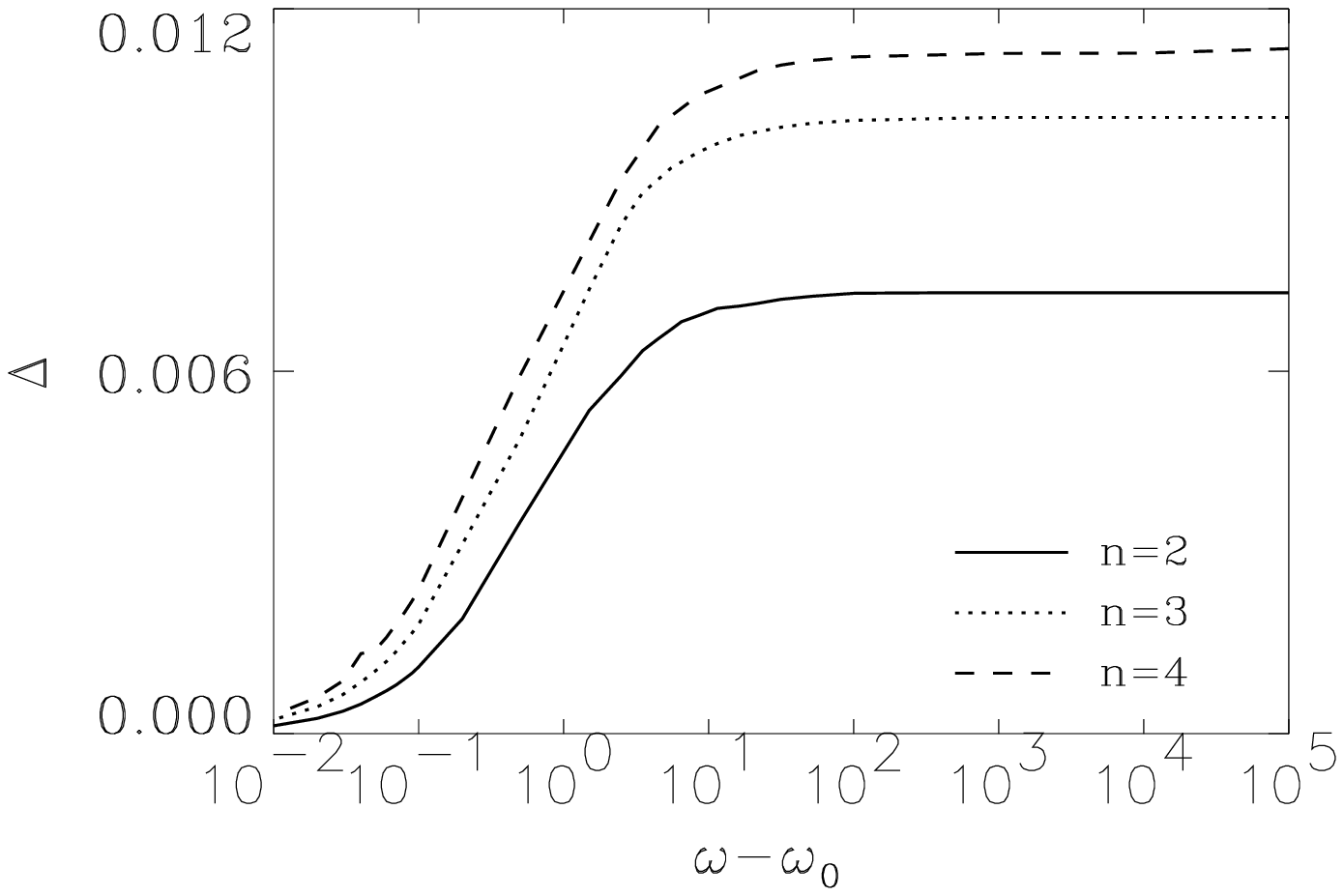}}
\caption{The difference $\Delta=E(n=1)-E(n)/n$ between 
the energy per winding number
of the $n=1$ monopole and the multimonopoles with winding number $n$
is shown as function of
$\omega-\omega_0$, $\omega_0=-\frac{3}{2}$, for $\alpha=1.0$, $\beta=0$ and
$n=2,3,4$, respectively.}
\end{figure}


\newpage
\begin{thebibliography}{000}
\bibitem{vilenkin} A. Vilenkin and E. P. S. Shellard. {\it Cosmic strings and other
topological defects}, Cambridge university press, 1994.
\bibitem{nielsen} H. B. Nielsen and P. Olesen, Nucl. Phys. {\bf B61} (1973), 45.
\bibitem{thooft} G. 'tHooft, Nucl. Phys. {\bf B79} (1974), 276;
A. M. Polyakov, JETP Lett. {\bf 20} (1974), 194. 
\bibitem{barriola} M. Barriola and A. Vilenkin, Phys. Rev. Lett. {\bf 63} (1989), 341.
\bibitem{harari} D. Harari and C. Lousto, Phys. Rev. {\bf D42} (1990), 2626. 
\bibitem{cohen} A. G. Cohen and D. B. Kaplan, Phys. Lett. {\bf B215} (1988), 67.
\bibitem{ruth} R. Gregory, Phys. Lett. {\bf B215} (1988), 663.
\bibitem{vilenkin2} A. Vilenkin, Phys. Rev. {\bf D23} (1981), 852.
\bibitem{bogo} B. Bogomol'nyi, Sov. J. Nucl. Phys. {\bf 24} (1976), 449;
D. Maison, Nucl. Phys. {\bf B182} (1981), 144.
\bibitem{rebbi} C. Rebbi and P. Rossi, Phys. Rev. {\bf D22} (1980), 2010.
\bibitem{bps} N. S. Manton, Nucl. Phys. {\bf B126} (1977), 525; E. J. Weinberg, 
Phys. Rev. {\bf D20} (1979), 936; J. N. Goldberg, P.S. Jang, S.Y. Park and K. C. Wali,
Phys. Rev. {\bf D16} (1978), 542; L. O'Raifeartaigh, S. Y. Park and K. Wali,
Phys. Rev. {\bf D20} (1979), 1941.
\bibitem{ps} M. K. Prasad and C. M. Sommerfield, Phys. Rev. Lett. {\bf 35} (1975)
449.
\bibitem{kkt} B. Kleihaus, J. Kunz and T. Tchrakian, Mod. Phys. Lett. {\bf A13}
(1998), 2523.
\bibitem{bfm} K. Lee, V. P. Nair and E. J. Weinberg, Phys. Rev. {\bf D45} (1992), 2751;
P. Breitenlohner, P. Forgacs and D. Maison, Nucl. Phys. {\bf B383} (1992), 357;
P. Breitenlohner, P. Forgacs and D. Maison, Nucl. Phys. {\bf B442} (1995), 126. 
\bibitem{hkk} B. Hartmann, B. Kleihaus and J. Kunz, Phys. Rev. Lett. {\bf 86} (2001),
1422.
\bibitem{bd} C. Brans and R. H. Dicke, Phys. Rev. {\bf 124} (1961), 925.
\bibitem{mach} E. Mach, {\it Conservation of energy}, Note No. 1 (1872);
{\it The Science of Mechanics} (1883), Chap. II, Sec. IV.
\bibitem{string} E. S. Fradkin and A. A. Tseytlin, Nucl. Phys. {\bf B261}
(1985), 1; 
C. G. Callan, D. Friedan, E.J. Martinec and M. J. Perry, Nucl. Phys. {\bf B262}
(1985), 593.
\bibitem{global} A. Barros and C. Romero, J. Math. Phys. {\bf 36} (1995), 5800;
Phys. Rev. {\bf D56} (1997), 6688; Phys. Rev. {\bf D60} (1999), 087502.
\bibitem{tmt} T. Tamaki, K. Maeda and T. Torii, Phys. Rev. {\bf D60} (1999), 104049.
\bibitem{bhk} Y. Brihaye, B. Hartmann and J. Kunz, Phys. Rev. {\bf D65} (2002), 024019.
\bibitem{bh} Y. Brihaye and B. Hartmann, Phys. Lett. {\bf B528} (2002), 288.
\bibitem{bhk} Y. Brihaye, B. Hartmann and J. Kunz, Phys. Rev. {\bf D65} (2002),
024019. 
\bibitem{bk2} B. Kleihaus and J. Kunz, Phys. Rev. Lett. {\bf 78} (1997), 2527.
\bibitem{theta} Since the solutions are axially symmetric, it is sufficient to integrate
the equations for $\theta\epsilon$ $[0:\pi/2]$.
\bibitem{z} This transformation maps the infinite interval $[0:\infty[$ to
the finite interval $[0:1]$.
\bibitem{bh_dila} Y. Brihaye and B. Hartmann, Phys. Lett. {\bf B528} (2002),
288.  
\bibitem{bh_volkov} Y. Brihaye and B. Hartmann, Phys. Lett. {\bf B534} (2002),
137.
\bibitem{volkov} M. S. Volkov, Phys. Lett. {\bf B524} (2002), 369.

\bibitem{hkk_bh} B. Hartmann, B. Kleihaus and J. Kunz, Phys. Rev. {\bf D65} (2002),
024027.
\bibitem{acs} A. Ashtekar, A. Corichi and D. Sudarsky, Class. Quantum Grav. 
{\bf 18} (2001) 919.
\end{thebibliography}
\end{document}